\documentclass[twocolumn,prd,letterpaper,twoside,preprintnumbers,showkeys]{revtex4}
\usepackage{graphicx}% Include figure files

\usepackage{fancyhdr}

\usepackage{pslatex}

\begin{document}

\title{Melting $\rho$ Meson and Thermal Dileptons}

\author{Ralf Rapp, Hendrik van Hees and Trenton Strong}

\affiliation{Cyclotron Institute and Physics Department, 
Texas A\&M University, College Station, TX 77843-3366, USA}

%\received{on ????????????, 2006}

\begin{abstract}
  We give a brief survey of theoretical evaluations of light vector
  mesons in hadronic matter, focusing on results from hadronic many-body
  theory. We emphasize the importance of imposing model constraints in
  obtaining reliable results for the in-medium spectral densities.  The
  latter are subsequently applied to the calculation of dilepton spectra
  in high-energy heavy-ion collisions, with comparisons to recent NA60
  data at the CERN-SPS. We discuss aspects of space-time evolution
  models and the decomposition of the excess spectra into different
  emission sources.


\end{abstract}
\maketitle

\thispagestyle{fancy}

\setcounter{page}{1}

%%%%%%%%%%%%%%%%%%%%%%%%%%
\section{INTRODUCTION}
%%%%%%%%%%%%%%%%%%%%%%%%%%
The investigation of hadron properties in strongly interacting matter
plays a central role for the understanding of the QCD phase diagram. On
the one hand, in-medium changes of hadronic spectral functions figure
into a realistic description of the bulk properties of hadronic matter,
i.e., its equation of state. On the other hand, hadronic modes
(especially those connected to order parameters) can serve as monitors
for approaching QCD phase transitions; in fact, hadronic correlations
may even play a significant role above $T_c$, e.g., as quark-antiquark
resonances with important consequences for transport
properties~\cite{vanHees:2005wb}.  Of particular interest are light
vector mesons ($V=\rho, \omega, \phi$), due to their decay channel into
dileptons whose invariant-mass spectra can provide undistorted
information on $V$-meson spectral functions in hot and/or dense hadronic
matter. A large excess of low-mass dielectrons measured in (semi-)
central Pb(158\,AGeV)+Au collisions at the CERN Super-Proton Synchrotron
(SPS)~\cite{ceres05} has established the presence of strong medium
effects in the electromagnetic spectral function. However, no definite
conclusion on the underlying mechanism of the excess could be drawn
(e.g., a dropping of the $\rho$ mass or a broadening of its
width)~\cite{Rapp:1999ej}.  More recently, the NA60 experiment measured
dimuon spectra in In(158\,AGeV)+In collisions~\cite{Arnaldi:2006jq} with
much improved statistics and mass resolution allowing for more stringent
conclusions: calculations based on hadronic many-body theory predicting
a strongly broadened $\rho$ spectral function~\cite{Rapp:1997fs} were
essentially confirmed, while those based on a dropping-mass
scenario~\cite{LKB96} are disfavored.

To evaluate consequences of medium-modified spectral densities for
experimental dilepton spectra at least two more ingredients are
required: (a) a space-time evolution model of the hot and dense system
in A-A reactions, which (in local thermal equilibrium) provides the
thermodynamic parameters (temperature, chemical potentials) for the
thermal emission rates, as well as blue shifts induced by collective
expansion, and (b) nonthermal sources, such as primordial Drell-Yan
annihilation, ``Corona'' effects or emission after thermal freezeout,
which become increasingly important at larger transverse momentum
($q_t$) and lower collision centralities.

In this contribution we further develop our 
interpretation~\cite{vanHees:2006ng} of the NA60 dilepton 
spectra~\cite{Arnaldi:2006jq}, including a more complete treatment 
of emission sources~\cite{vanHees:2006ng} and $q_t$ dependencies.
We briefly review calculations of in-medium low-mass vector spectral 
functions in Sec.~\ref{sec_vmes} and discuss their application
to NA60 data in Sec.~\ref{sec_dilep}, including recently published
$q_t$-dependencies.  Sec.~\ref{sec_concl} contains our conclusions. 
%In particular, we discuss recently published
%transverse-momentum dependencies and  possible interpretations  
%in terms of underlying emission sources.

%%%%%%%%%%%%%%%%%%%%%%%%%%%%%%%%%%%%%%%%%%%%%
\section{VECTOR MESONS IN HADRONIC MATTER}
\label{sec_vmes}
%%%%%%%%%%%%%%%%%%%%%%%%%%%%%%%%%%%%%%%%%%%%%
%\subsection{Spectral Functions}
Hadronic approaches to calculate in-medium vector-meson spectral 
functions are typically based on effective Lagrangians with parameters 
(masses, coupling constants and vertex formfactors) adjusted 
to empirical decay rates (both hadronic and electromagnetic)
and scattering data (e.g., $\pi N\to VN$ or nuclear photoabsorption). 
The interaction vertices are implemented into a hadronic many-body 
scheme to calculate selfenergy insertions 
of the $\rho$-propagator in hot/dense matter, 
\begin{equation}
D_V(M,q;\varrho_B,T)=[M^2-m_V^2-\Sigma_{VP}
                     -\Sigma_{VB} -\Sigma_{VM} ]^{-1}
\label{prop}
\end{equation}  
($T$: temperature, $\varrho_B$: baryon density); $\Sigma_{VM,VB}$
accounts for direct interactions of $V$ with surrounding mesons and
baryons, and $\Sigma_{VP}$ for the in-medium pseudoscalar meson cloud
($2\pi$, $3\pi$ and $K\bar K$ for $V$=$\rho$, $\omega$ and $\phi$,
respectively).

\begin{figure}[!b]
\vspace*{-0.10cm}
\begin{center}
\includegraphics*[width=5.5cm,angle=-90]{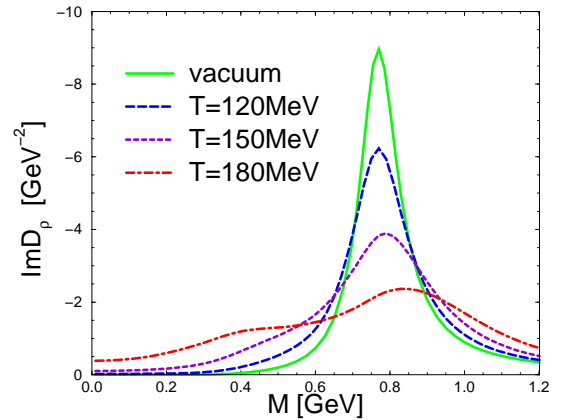}
\end{center}
\vspace*{-0.4cm}
\caption{$\rho$-meson spectral function, ${\rm Im}D_\rho(M,q=0)$, in
  hadronic matter under approximate SPS conditions ($\mu_B$=330\,MeV).}
\label{fig_rho}
\end{figure}
Due to its prime importance for low-mass dilepton emission
(cf.~eq.~(\ref{piem}) below), many theoretical studies have focused on
the $\rho$ meson. A typical result for its spectral
function~\cite{Rapp:1997fs} is shown in Fig.~\ref{fig_rho}, indicating a
strong broadening with increasing $T$ and $\varrho_B$ and little mass
shift.

An important question is whether the parameters of the effective
Lagrangian (e.g., the bare mass, $m_V$, in eq.~(\ref{prop})) depend on
$T$ and $\varrho_B$. This requires information beyond the
effective-theory level.  In Ref.~\cite{Harada:2003jx}, a
Hidden-Local-Symmetry framework for introducing the $\rho$ meson into a
chiral Lagrangian has been treated within a renormalization group
approach at finite temperature (pion gas).  While the hadronic
interactions affect the in-medium $\rho$ properties only moderately, a
matching of the vector and axialvector correlator to the (spacelike)
operator product expansion (involving quark and gluon condensates which
decrease with increasing $T$) requires a $T$-dependence of hadronic
couplings and bare masses, inducing a dropping $\rho$-mass.  On the
other hand, based on QCD sum rules in cold nuclear matter (which also
involve an operator product expansion), it was shown that the decrease
of in-medium quark and gluon condensates can also be satisfied by an
increased width of the $\rho$ spectral function~\cite{Leupold:1997dg}.
The finite-density spectral functions corresponding to
Fig.~\ref{fig_rho} are compatible with the QCD sum rule constraints of
Ref.~\cite{Leupold:1997dg}. Since the latter are mostly driven by
in-medium 4-quark condensates, $\langle (q\bar q)^2\rangle$, a more
accurate determination of these is mandatory to obtain more stringent
conclusions on the $V$ spectral functions~\cite{Thomas:2005dc}.

For the $\omega$ meson, hadronic Lagrangian calculations predict an
appreciable broadening as well ($\Gamma_\omega^{\rm med}$$\simeq$50\,MeV
at normal nuclear matter density, $\varrho_0$=0.16\,fm$^{-3}$), but
reduced masses have also been found~\cite{Klingl:1998zj}.  A recent
experiment on photoproduction of $\omega$ mesons off
nuclei~\cite{CBTAPS} has provided evidence for a decreased $\omega$
mass, but an interpretation of the data using an $\omega$ spectral
function with 90~MeV width seems also viable~\cite{Kaskulov:2006zc}.

Finally, $\phi$ mesons are expected to undergo significant broadening in
hot and dense matter~\cite{Alvarez-Ruso:2002ib,Cabrera:2003wb}, mostly
due to modifications of its kaon cloud. Nuclear photoproduction of
$\phi$ mesons~\cite{Ishikawa:2004id} indicates absorption cross sections
that translate into an in-medium width of $\sim$50\,MeV at $\varrho_0$.

%%%%%%%%%%%%%%%%%%%%%%%%%%%%%%%%%%%%%%%
\section{DILEPTON SPECTRA AT CERN-SPS}
\label{sec_dilep}
%%%%%%%%%%%%%%%%%%%%%%%%%%%%%%%%%%%%%%%
In a hot and dense medium, the equilibrium emission rate of dileptons
($l^+l^-$ with $l$=$e$,$\mu$) can be written as~\cite{MT84}
\begin{equation}
\frac{d N_{ll}}{d^4 x d^4 q}
=-\frac{\alpha^2}{\pi^3} \frac{L(M^2)}{M^2} \ 
{\rm Im}\Pi_{\rm em}(M,q) \ f^B(q_0;T)
\label{rate}
\end{equation}
($f^B$: Bose distribution, $L(M^2)$: lepton phase space,
$\alpha=1/137$). Within the vector dominance model, the low-mass regime
($M$$\le$1\,GeV) of the retarded e.m.~current-current correlator can be
saturated by the light vector mesons,
\begin{equation}
{\rm Im}\Pi_{\rm em} = \sum\limits_{V=\rho,\omega,\phi} 
\frac{m_V^4}{g_V^2} \mathrm{Im}D_V \ , 
\label{piem}
\end{equation} 
implying an approximate weighting of $\rho$:$\omega$:$\phi$
contributions of 10:1:2 (reflecting the values of $\Gamma_{V\to ee}$).

\begin{figure}[!t]
\vspace*{0.3cm}
\begin{center}
\hspace{-0.2cm}
\includegraphics*[width=7.1cm,angle=0]{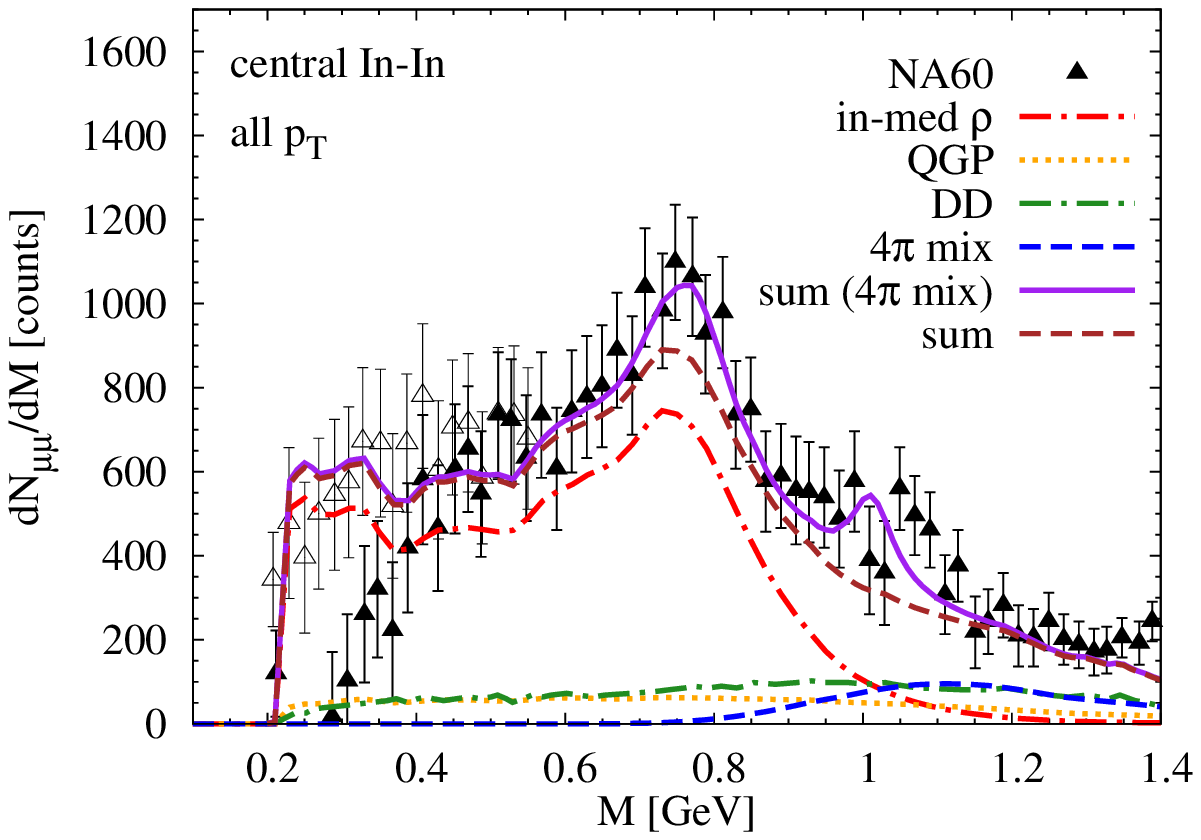}
\includegraphics*[width=7.1cm,angle=0]{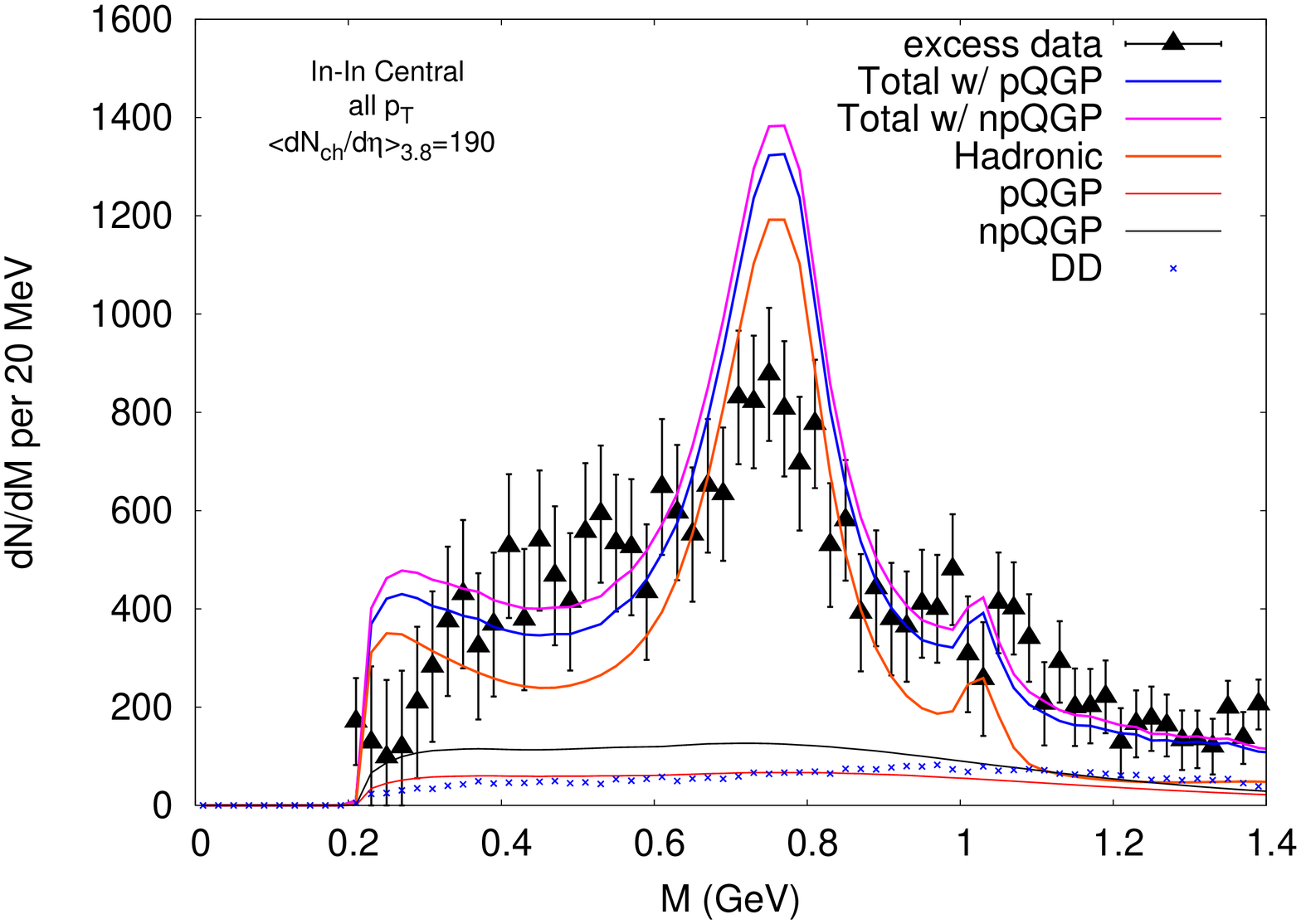}
\end{center}
\vspace*{-0.4cm}
\caption{Excess dilepton mass spectra in central In(158\,AGeV)-In
  compared to (a) thermal fireball calculations with thermal emission
  from QGP, in-medium $\rho$, $\omega$ and $\phi$ mesons and 4-pion-type
  annihilation~\cite{vanHees:2006ng} (upper panel) (b) a hydrodynamic
  convolution of rates from the chiral virial
  approach~\cite{Dusling:2006yv} (lower panel).}
\label{fig_na60-cent}
\end{figure}
\begin{figure}[!b]
\vspace*{0.0cm}
\begin{center}
\includegraphics*[width=7.0cm,angle=-0]{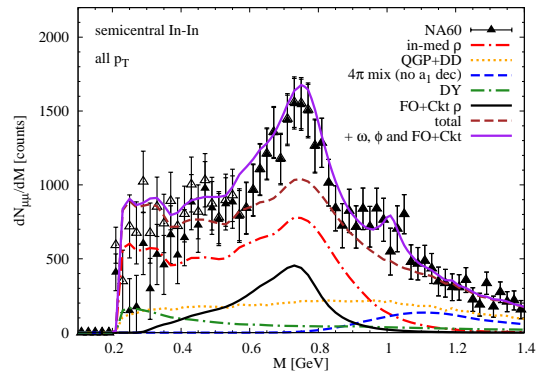}
\end{center}
\vspace*{-0.4cm}
\caption{Same as Fig.~\ref{fig_na60-cent} (upper panel), but for
  semicentral In(158\,AGeV)-In; the calculations additionally include
  primordial Drell-Yan annihilation and $\rho$ decays after thermal
  freezeout.}
\label{fig_na60-semi}
\end{figure}
Thermal dilepton spectra in A-A collisions are obtained by convoluting
the rate (\ref{rate}) over space-time. We employ an expanding thermal
fireball whose parameters are tuned to hydrodynamic simulations (see,
e.g., Ref.~\cite{Kolb:2003dz}). For central In(158\,AGeV)-In, an initial
Quark-Gluon Plasma (QGP) phase ($T_0$=197\,MeV) is followed by a mixed
phase and hadrochemical freezeout at ($\mu_B^c,T_c$)=(232,175)\,MeV. The
subsequent hadronic phase incorporates meson chemical potentials to
conserve the measured particle ratios, with thermal freezeout at $T_{\rm
  fo}$$\simeq$120\,MeV after a total lifetime of $\sim$7\,fm/c.  In the
QGP, dilepton emission is due to $q\bar q$ annihilation, while in the
hadron gas it is governed by in-medium $\rho$, $\omega$ and $\phi$ at
low mass and 4$\pi$-type annihilation at $M$$\ge$1\,GeV (as inferred
from the vacuum e.m.~correlator)~\cite{vanHees:2006ng}. The calculated
spectra describe the NA60 excess data well (upper panel of
Fig.~\ref{fig_na60-cent}).  This implies a $\rho$ spectral function that
has essentially ``melted'' around $T_c$$\simeq$175\,MeV (while there is
currently little sensitivity to in-medium $\omega$ and $\phi$ line
shapes).  The lower panel of Fig.~\ref{fig_na60-cent} compares the NA60
data to thermal emission spectra based on the chiral virial approach
when folded over a hydrodynamic evolution~\cite{Dusling:2006yv} (the
results agree well with our fireball convolution using the same input
rates~\cite{vanHees:2006iv}).  Albeit the virial rates imply a quenching
of the $\rho$ peak, a lack of broadening is apparent. Below the free
$\rho$ mass, the emission strength in the virial expansion is similar to
the in-medium $\rho$ spectral function.

\begin{figure}[!tb]
\vspace*{0.3cm}
\begin{center}
\includegraphics*[width=7.0cm,angle=-0]{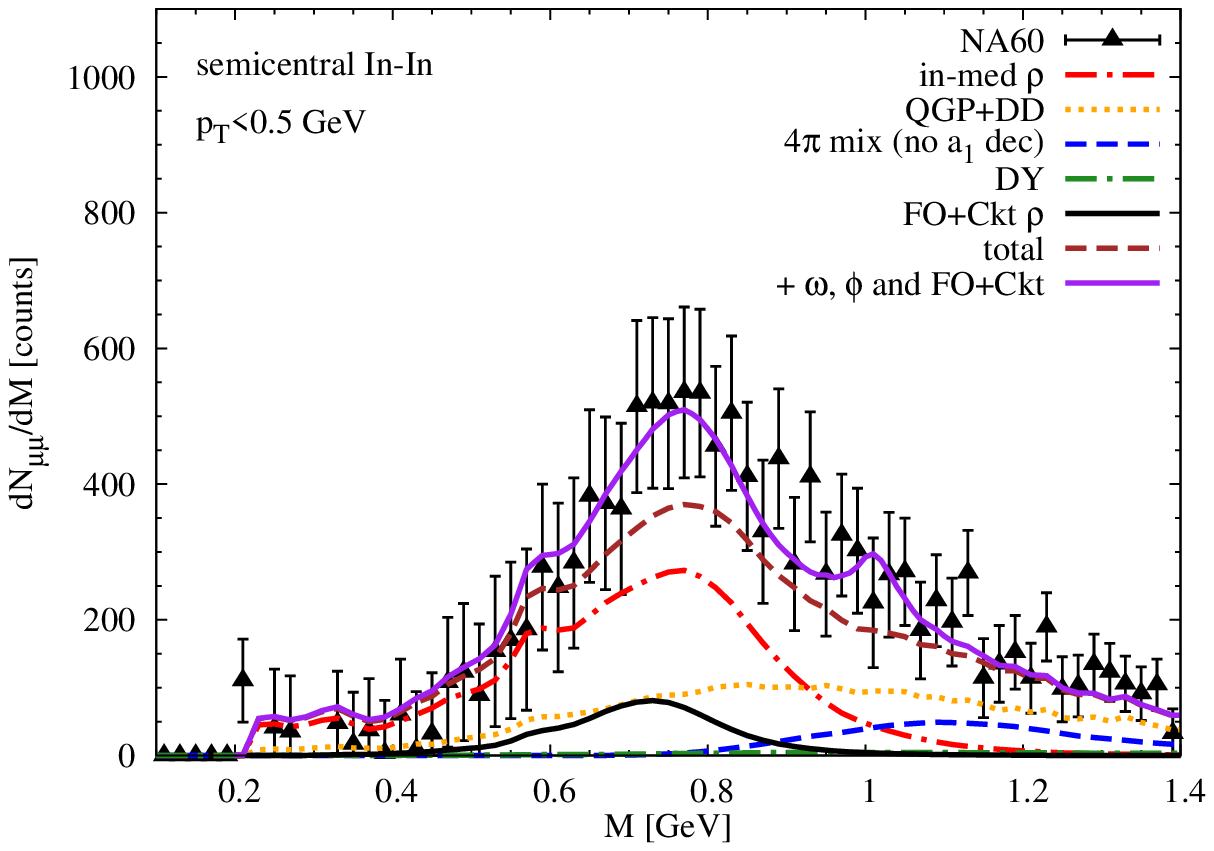}
\vspace*{0.3cm}
\includegraphics*[width=7.0cm,angle=-0]{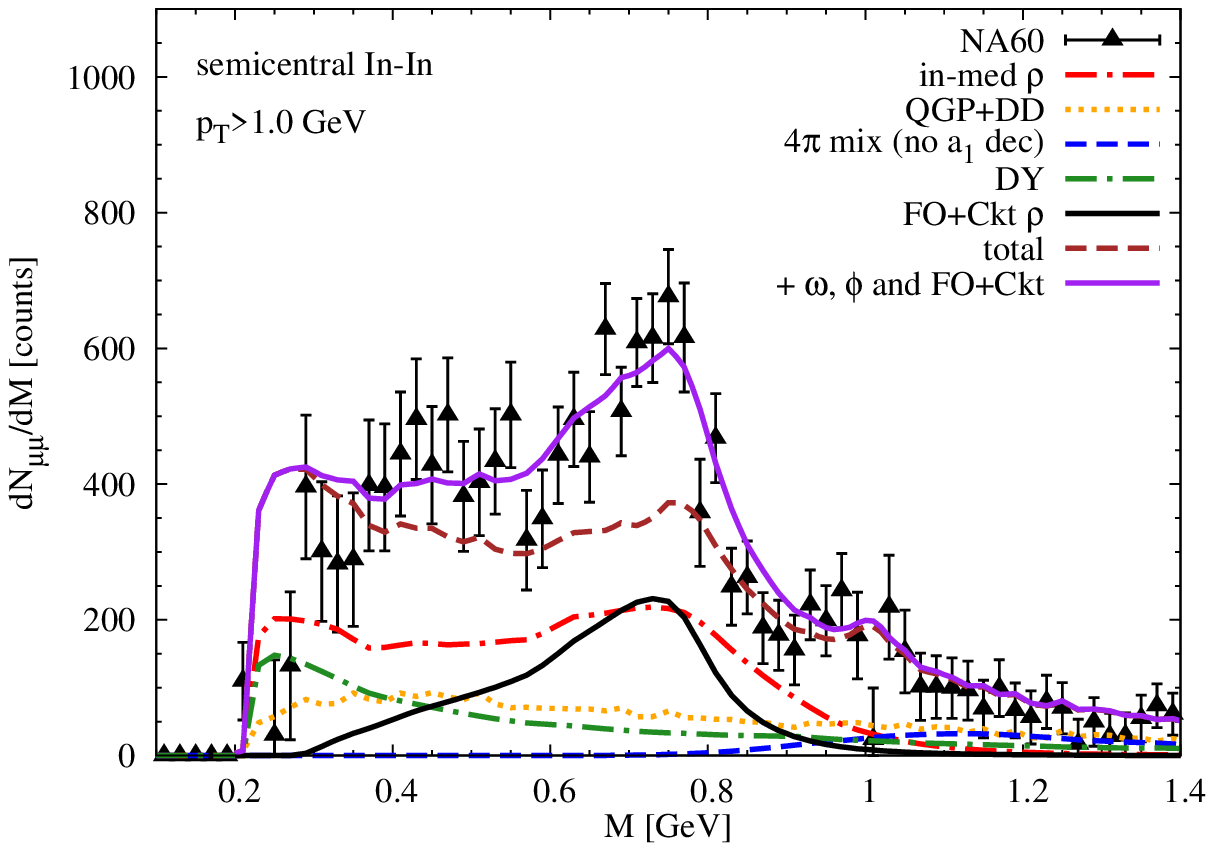}
\end{center}
\vspace*{-0.4cm}
\caption{Same as Fig.~\ref{fig_na60-semi}, but sliced into 2
bins of transverse pair momentum: $q_t$$<$$0.5$\,GeV (upper panel)
and $q_t$$>$$1.0$\,GeV (lower panel).}
\label{fig_na60-semi2}
\end{figure}
For a more accurate treatment of $\rho$ decays at thermal freezeout, the
rate formula (\ref{rate}) should be replaced by a Cooper-Frye type
hydrodynamic freezeout~\cite{Cooper:1974mv,Skokov:2005ut,Renk:2006ax}
which leads to somewhat modified kinematics. Furthermore, the spectral
function after freezeout is not expected to carry the full medium
effects anymore.  In Fig.~\ref{fig_na60-semi} we implement these
improvements, along with primordial Drell-Yan annihilation (DY) and
initial ``corona'' $\rho$'s, and compare to semicentral NA60 data. Since
the new contributions are characterized by rather hard $q_t$ slopes
(large ``effective temperatures''), their significance primarily resides
at high $q_t$.  This is borne out when dividing the mass spectra in
$q_t$ bins~\cite{Damjanovic:2006bd}: freezeout/corona $\rho$'s and DY
mostly contribute at high $q_t$ leading to fair agreement with
experiment, cf.~Fig.~\ref{fig_na60-semi2}.

\vspace*{3mm}
\section{CONCLUSIONS}
\label{sec_concl}

Hadronic many-body calculations predict a strong broadening, and little
mass shift, of the $\rho$-meson spectral function in hot/dense hadronic
matter. The good agreement with improved dilepton invariant-mass spectra
by NA60 at the SPS supports the notion of a ``melting'' $\rho$-meson
close to the expected phase boundary to the QGP. The next challenge is
to establish rigorous connections to the properties of thermal QCD,
especially to chiral symmetry restoration. There are promising prospects
that, using chiral effective models and constraints from lattice QCD,
this goal can be achieved by systematic (and quantitative) evaluations
of Weinberg and QCD sum rules, which directly relate axial-/vector
spectral functions to chiral order parameters.

\vspace*{0.28cm}

\noindent{\bf Acknowledgments} 
This work has been supported by a U.S. National Science 
Foundation (NSF) CAREER Award under grant PHY-0449489, and by 
the NSF REU program under grant no. PHY-0354098.

%\smallskip 


\begin{thebibliography}{99}

\bibitem{vanHees:2005wb}
H.~van Hees, V.~Greco, R.~Rapp, 
Phys. Rev. C {\bf 73} (2006) 034913.

\bibitem{ceres05}
CERES/NA45 Collaboration [G. Agakichiev {\it et al.}], 
Eur. Phys. J. C {\bf 41} (2005) 475.

\bibitem{Rapp:1999ej}
R.~Rapp and J.~Wambach,
%   ``Chiral symmetry restoration and dileptons in relativistic heavy-ion
%collisions,''
Adv.\ Nucl.\ Phys.\  {\bf 25} (2000) 1.
%  [arXiv:hep-ph/9909229].
%%CITATION = HEP-PH 9909229;%%

\bibitem{Arnaldi:2006jq}
 NA60 Collaboration [R.~Arnaldi {\it et al.}],
%``First measurement of the rho spectral function in high-energy nuclear
%collisions,''
  Phys.\ Rev.\ Lett.\  {\bf 96} (2006) 162302.
%  [arXiv:nucl-ex/0605007].
%%CITATION = NUCL-EX 0605007;%%

\bibitem{Rapp:1997fs}
R.~Rapp, G.~Chanfray, J.~Wambach,
%``Rho meson propagation and dilepton enhancement in hot hadronic matter,''
Nucl.\ Phys.\ {\bf A617} (1997) 472.
%[arXiv:hep-ph/9702210].
%%CITATION = HEP-PH 9702210;%%

\bibitem{LKB96}
G.Q.~Li, C.M.~Ko, G.E.~Brown, Nucl. Phys. {\bf A606} (1996) 568.

\bibitem{vanHees:2006ng}
  H.~van Hees and R.~Rapp,
  %``Comprehensive interpretation of thermal dileptons at the SPS,''
  Phys. Rev. Lett. {\bf 97} (2006) 102301.
  %%CITATION = HEP-PH 0603084;%%

\bibitem{Harada:2003jx}
  M.~Harada and K.~Yamawaki,
%``Hidden local symmetry at loop: A new perspective of composite gauge boson
%and chiral phase transition,''
  Phys.\ Rept.\  {\bf 381} (2003) 1.
%  [arXiv:hep-ph/0302103].
  %%CITATION = HEP-PH 0302103;%%
                                                                                
\bibitem{Leupold:1997dg}
  S.~Leupold, W.~Peters, U.~Mosel,
  %``What QCD sum rules tell about the rho meson,''
  Nucl.\ Phys.\ {\bf A628} (1998) 311.
%  [arXiv:nucl-th/9708016].
%%CITATION = NUCL-TH 9708016;%%

\bibitem{Thomas:2005dc}
  R.~Thomas, S.~Zschocke, B.~K\"ampfer,
  %``Evidence for in-medium changes of four-quark condensates,''
  Phys.\ Rev.\ Lett.\  {\bf 95} (2005) 232301.
%  [arXiv:hep-ph/0510156].
%%CITATION = HEP-PH 0510156;%%

\bibitem{Klingl:1998zj}
  F.~Klingl, T.~Waas, W.~Weise,
  %``Nuclear bound states of omega mesons,''
  Nucl.\ Phys.\ A {\bf 650} (1999) 299.
%  [arXiv:hep-ph/9810312].
%%CITATION = HEP-PH 9810312;%%
                                                                                
\bibitem{CBTAPS}
CBELSA/TAPS Collaboration [D.~Trnka {\it et al.}],
Phys. Rev. Lett. {\bf 94} (2005) 192303.
                                                                                
\bibitem{Kaskulov:2006zc}
  M.~Kaskulov, E.~Hernandez, E.~Oset,
  %``Inclusive omega photoproduction from nuclei and omega in the nuclear
  %medium,''
  arXiv:nucl-th/0610067.
  %%CITATION = NUCL-TH 0610067;%%

\bibitem{Alvarez-Ruso:2002ib}
L.~Alvarez-Ruso and V.~Koch, Phys. Rev. C \textbf{65} (2002) 054901.
                                                                                
\bibitem{Cabrera:2003wb}
  D.~Cabrera {\it et al.}, 
%L.~Roca, E.~Oset, H.~Toki, M.J.~Vicente Vacas,
  %``Mass dependence of inclusive nuclear Phi photoproduction,''
  Nucl.\ Phys.\ {\bf A733} (2004) 130.
%  [arXiv:nucl-th/0310054].
%%CITATION = NUCL-TH 0310054;%%

\bibitem{Ishikawa:2004id}
  T.~Ishikawa {\it et al.},
%``Phi photo-production from Li, C, Al, and Cu nuclei at E(gamma) =  1.5-GeV -
%2.4-GeV,''
Phys.\ Lett.\ B {\bf 608} (2005) 215.
%[arXiv:nucl-ex/0411016].
%%CITATION = NUCL-EX 0411016;%%

\bibitem{MT84}
L.D.~McLerran and T.~Toimela, Phys. Rev. D \textbf{31} (1985) 545.

\bibitem{Kolb:2003dz}
P.F.~Kolb and U.W.~Heinz,
%``Hydrodynamic description of ultrarelativistic heavy-ion collisions,''
arXiv:nucl-th/0305084.
%%CITATION = NUCL-TH 0305084;%%

\bibitem{Dusling:2006yv}
  K.~Dusling, D.~Teaney, I.~Zahed,
  %``Thermal dimuon yields at NA60,''
  arXiv:nucl-th/0604071.
  %%CITATION = NUCL-TH 0604071;%%

\bibitem{vanHees:2006iv}
  H.~van Hees and R.~Rapp,
  %``Medium modifications of vector mesons and NA60,''
  arXiv:hep-ph/0604269.
  %%CITATION = HEP-PH 0604269;%%

\bibitem{Cooper:1974mv}
  F.~Cooper and G.~Frye,
%``Comment On The Single Particle Distribution In The Hydrodynamic And
%Statistical Thermodynamic Models Of Multiparticle Production,''
  Phys.\ Rev.\ D {\bf 10} (1974) 186.
  %%CITATION = PHRVA,D10,186;%%                                      

\bibitem{Skokov:2005ut}
  V.~V.~Skokov and V.~D.~Toneev,
  %``Semi-central In-In collisions and Brown-Rho scaling,''
  Phys.\ Rev.\ C {\bf 73} (2006) 021902.
%  [arXiv:nucl-th/0509085].
  %%CITATION = NUCL-TH 0509085;%%

\bibitem{Renk:2006ax}
  T.~Renk and J.~Ruppert,
  %``What the NA60 dilepton data can tell,''
  arXiv:hep-ph/0605130.
  %%CITATION = HEP-PH 0605130;%%


\bibitem{Damjanovic:2006bd}
S.~Damjanovic [NA60 Collaboration],
%``First measurement of the rho spectral function in nuclear collisions,''
arXiv:nucl-ex/0609026.
%%CITATION = NUCL-EX 0609026;%%

%\bibitem{HSR06}
%H.~van Hees, T.~Strong and R.~Rapp, in preparation.


\end{thebibliography}
\end{document}